\documentclass{elsart3}

\usepackage{mathptmx}
\usepackage[dvips]{graphicx}

\usepackage{amssymb}

\begin{document}

\begin{frontmatter}



\title{Structure and magnetism in nanocrystalline Ca(La)B$_6$ films}


\author[apph] {Y. Sakuraba},
\author[apph]{H. Kato\corauthref{cor1}},
\ead{kato@mlab.apph.tohoku.ac.jp}
\author[apph]{F. Sato},
\author[apph] {T. Miyazaki},
\author[clts] {N. Kimura},
\author[clts] {H. Aoki}
\address[apph]{Department of Applied Physics, Tohoku University,
       Aoba-yama 05, Sendai 980-8579, Japan}
\address[clts]{Center for
Low Temperature Science, Tohoku University., Aoba-ku, Sendai 980-8578, Japan}       
\corauth[cor1]{Corresponding author.  Fax: +81-22-217-7947.}

\begin{abstract}

Nanocrystalline films of La-doped CaB$_6$  have been fabricated by using a rf-magnetron sputtering. Lattice 
expansion of up to 6\% with respect to the bulk value
was observed along the direction perpendicular to the film plane, which arises from the trapping of Ar gas into the film.   Large ferromagnetic moment of $3\sim4$  $\mu_{\rm B}$/La has been observed in some La-doped films only when the lattice expansion rate is larger than 2.5\%.
\end{abstract}

\end{frontmatter}


\label{}The reported magnetism in lanthanum-doped CaB$_6$ \cite{young} has attracted great interest, because of its high Curie temperature $T_{\rm C}$ in spite of having no 3$d$ or 4$f$ element as a constituent.  It has been reported, however, there is a large sample dependence
of magnetic and electric properties even for carefully-prepared single crystals.
Matsubayashi {\it et al.} \cite{matsubayashi} claimed that the magnetism observed in this system is not intrinsic but due to the alien phases of iron and boride.
So the origin of the ferromagnetism remains  controversial.  In this paper we briefly report the structure and magnetism in sputter-deposited nanocrystalline films of La-doped CaB$_6$.

Thin films of Ca$_{0.995}$La$_{0.005}$B$_6$ were fabricated by using an rf-magnetron sputtering method.  Either Ar or Ne was used as a sputtering gas.  The Si (100) substrate was heated during the deposition up to $T_{\rm s} = 900^\circ$C.   In addition to the conventional X-ray diffraction (XRD) measurements in $\theta$-$2\theta$ mode which give an information about the lattice spacing perpendicular to the film plane, we also examined an in-plane XRD pattern by using RIGAKU ATX-G system, in which the scattering vector is almost parallel to the film plane.

A typical example of XRD patterns for the films deposited by using the Ar gas is shown in Fig. 1(a).  As for the pattern taken by $\theta$-$2\theta$ scans,  Bragg peak positions are
  shifted significantly toward the lower-angle side with respect to those for bulk CaB$_6$. 
This suggests a lattice expansion along the direction perpendicular to the film plane, with respect to the bulk value of $a_{\rm bulk}=4.145$ \AA.
We deduced a hypothetical cell parameter $a_{\perp}$ according to the position of the (110) peak, which was found to decrease with  increasing Ar gas pressure $P_{\rm Ar}$ and film thickness $t$.  However,  it is still about 1\% larger than the bulk value for the sample with 
$t=8$  $\mu$m, $P_{\rm Ar}=5$ mTorr, and $T_{\rm s}=700^\circ$C.  
Also plotted in Fig. 1(a) is a XRD pattern for the same film but measured by the in-plane mode, in which the peak positions are slightly higher than those for the bulk sample. This indicates that the spacing along the directions within the film plane is slightly shrunk ($a_{//} < a_{\rm bulk}$), in striking contrast with that $a_{\perp} > a_{\rm bulk}$.  Average grain sizes estimated from a width of the diffraction peaks are also anisotropic with $d_{//}\sim  10$ nm and $d_{\perp}\sim 30$ nm, respectively. 
  
\begin{figure}[t]
\begin{center}
\includegraphics[width=7.8cm,clip]{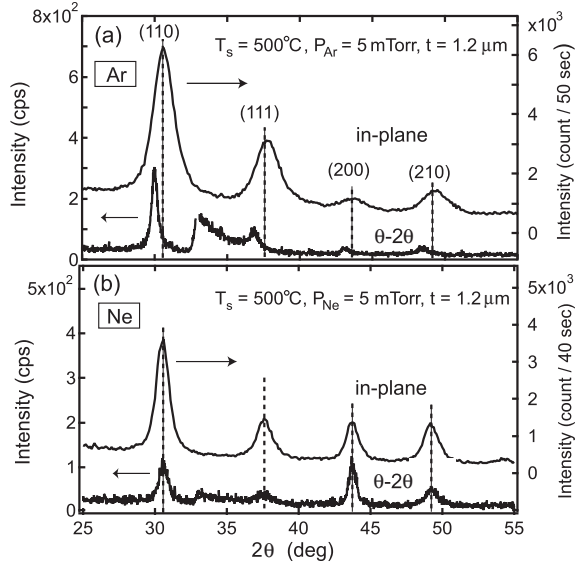}
\end{center}
\caption{XRD patterns for Ca$_{0.995}$La$_{0.005}$B$_6$ films sputtered by using Ar (a) and Ne (b) gases. The upper and lower patterns in each graph indicate the data measured by in-plane mode and 
by normal $\theta$-$2\theta$ scan, respectively.
The vertical broken lines indicate the peak positions of bulk CaB$_6$. }
\label{fig1}
\end{figure}

The SEM-EDS analysis have revealed that a significant amount of Ar has been 
trapped into a film during a sputtering.  We plotted in Fig. 2(a) the Ar content as functions of $a_{//}$ and $a_{\perp}$ for a large number of films. 
It is clearly seen that, with increasing Ar content of up to about 0.5 at\%,  $a_{\perp}$ increases (up to 6\%), while $a_{//}$ decreases (up to about 1\%) with respect to $a_{\rm bulk}$.
As for the Ne-sputtered films, no significant deviation of $a_{//}$ and $a_{\perp}$ from the $a_{\rm bulk}$ was observed as shown in Fig. 1(b) and 2(b), although the trapped Ne as high as 0.15 at\% was detected.

Magnetization isotherms of these films were measured by using the SQUID magnetometer, some of which exhibit ferromagnetic-like hysteresis loops with the coercive field of about 70 Oe at room temperature.   Spontaneous magnetic moment $M_{\rm s}$ was deduced by the linear extrapolation of $M$ vs $H$ curve in the high-field region to zero field. 
It was found that $M_{\rm s}(T=400$ K) is about 80\% of $M_{\rm s}(T=5$ K), suggesting that $T_{\rm C}$ is significantly higher than 400 K.
The largest $M_{\rm s}$ value at room temperature amounts to 4 $\mu_{\rm B}$ per La atom, assuming the nominal concentration of La (0.5 at\%).  We plotted the $M_{\rm s}$ value thus deduced in Fig. 2(b), as a function of $a_{\perp}$.  
Although the data points are scattered, there is a tendency for $M_{\rm s}$ to increase with  increasing $a_{\perp}$.
Present results appear to be consistent with theoretical model proposed by Hotta {\it et al.} \cite{Hotta},  in which ferromagnetism is stabilized if the symmetry of the system is tetragonal rather than cubic.
Experimentally, there would be a correlation with the defect-driven-enhancement phenomena in bulk CaB$_6$ \cite{Lofland}.

\begin{figure}[t]
\begin{center}
\includegraphics[width=6.0cm,clip]{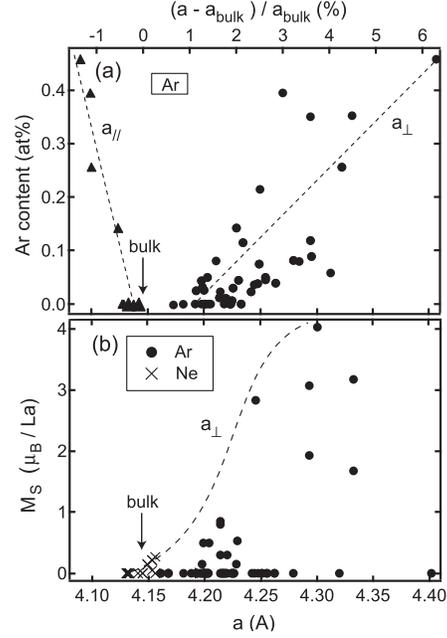}
\end{center}
\caption{Trapped Ar content plotted against hypothetical cell parameters $a_{//}$ and
 $a_{\perp}$ (a), and spontaneous magnetic moment per La as a function
 of  $a_{\perp}$ for both Ar- and Ne-sputtered films (b).  
  }
\label{fig2}
\end{figure}




\end{document}